\documentclass[aps,prb,twocolumn,showpacs]{revtex4-2}

\usepackage{hyperref}
\usepackage[dvipdfmx]{graphicx}
\graphicspath{{figures/}}
\usepackage{amsmath}
\usepackage{amsfonts}
\usepackage{xcolor}
\usepackage{braket}
\usepackage[normalem]{ulem}
\usepackage{listings}

\hypersetup{colorlinks=true, citecolor=blue, urlcolor=blue, linkcolor=blue}

\newcommand{\s}{\mathbf{S}}

\usepackage{xcolor}
\definecolor{orange}{rgb}{1,0.5,0}
\definecolor{darkred}{rgb}{0.55,0,0}

\usepackage[normalem]{ulem}
\usepackage{comment}

\newcommand{\yyy}[1]{ { \color{blue} \footnotesize (\textsf{yyy}) }}

\begin{document}
\title{
Chiral Spin Liquid on a Shastry-Sutherland Heisenberg Antiferromagnet
}

\author{Jian-Wei Yang$^{1,*}$, Wei-Wei Luo$^{2,3,*}$, W. Zhu$^{2,3}$, L. Wang$^4$, Bo Yang$^{5}$, Pinaki Sengupta$^5$}

\affiliation{$^1$School of Microelectronics and Data Science, Anhui University of Technology, Maanshan 243002, China}
\affiliation{$^2$Institute of Natural Sciences, Westlake Institute for Advanced Study, Hangzhou 310024, China}
\affiliation{$^3$School of Science, Westlake University, Hangzhou 310024, China}
\affiliation{$^4$ Department of Physics, Zhejiang University, Hangzhou 310000, China}
\affiliation{$^5$ School of Physical and Mathematial Sciences, Nanyang Technological University, Singapore 637371 }

\date{\today}

\begin{abstract}

    We demonstrate the existence of a topological chiral spin liquid in the frustrated Shastry-Sutherland Heisenberg model with an additional spin chirality interaction, using numerically unbiased exact diagonalization and density matrix renormalization group methods. We establish a quantum phase diagram where conventional phases, including dimer singlet, plaquette singlet, N{\' e}el and collinear phase, can be clearly identified by suitable local order parameters. Among them a $SU(2)_1$ chiral spin liquid emerges in the highly frustrated region, which is unambiguously identified by two topologically degenerate ground states, modular matrix, and characteristic level counting in entanglement spectrum, featuring the same topological order of $\nu=1/2$ bosonic Laughlin state.
    The phase boundaries among the different orders are determined by the energy level crossing analysis and wave function fidelity susceptibility.
\end{abstract}

\maketitle

\section{Introduction}

As one of the most intriguing quantum phases in condensed matter physics, quantum spin liquid (QSL)~\cite{balents2010,zhou2017,savary2017,broholm2020} does not form any conventional order even down to zero temperature. Consequently, such quantum state of matter goes beyond the description of Landau's symmetry breaking paradigm. Interestingly, the quantum disordered QSL has a rich organizing pattern internally and possesses fractionalized quasi-particles excitations and long range quantum entanglement, which keeps attracting great interests in the community since the initial proposal by Anderson 50 years ago~\cite{anderson1973}. The chiral spin liquid (CSL)~\cite{kalmeyer1987}, a special type of gapped QSL that breaks time reversal symmetry, is closely related to the fractional quantum Hall liquid~\cite{klitzing1980,tsui1982,laughlin1983} and thus also exhibits nontrivial topological order~\cite{wen1990}.
For the fractional quantum Hall system, gapped ground states have a topological degeneracy that depends on the lattice geometry, quasiparticle excitations possess fractional statistics, and gapless edge excitations exhibit a characteristic level counting that manifests underlying topological order in the bulk. These unique features can be used to identify topological ordered phases including CSL.
Over the last decade, the CSL has been unambiguously demonstrated in various lattice spin models by large scale numerical methods, including those on kagome lattice~\cite{gong2014,he2014a,bauer2014,he2015b}, triangular lattice~\cite{gong2017,cookmeyer2021}, honeycomb lattice~\cite{hickey2016,hickey2017,huang2021}, and square lattice~\cite{nielsen2013,hickey2017,merino2022}. Quite recently, emergent CSL has also been reported in the half-filled triangular Hubbard model~\cite{szasz2020,chen2022a}, sandwiched between the metallic and Mott insulating phases. Doping such state may lead to superconductivity.
The mechanism of the formation of the CSL is attributed to the strong interplay of geometric frustration and quantum fluctuation, which serves as a guiding principle to search CSL in realistic models and materials.

Besides the aforementioned widely studied lattice models, another lattice system with intrinsic frustration is the Heisenberg antiferromagnetic model on the Shastry-Sutherland (SS) lattice~\cite{shastry1981} with inter-dimer $J$ and intra-dimer $J^{\prime}$ interactions (see inset of Fig.\ref{pd}). 
The SS lattice can be realized by the compound 
SrCu$_2$(BO$_3$)$_2$ \cite{SCBO1991,SCBO1999}, where 
the in-plane spin-$1/2$ Cu spins are coupled by Heisenberg interactions. Interestingly, the relative strengths of these two couplings can be tuned by applying suitable pressure in experiment. 
Under ambient pressure intra-dimer interaction dominates and a dimer singlet (DS) phase is realized~\cite{SCBO1999,Miyahara1999}. The N{\' e}el order has also been observed at high pressure, while some variant of plaquette singlet (PS) phase is detected around 2GPa~\cite{zayed2017,guo2020,bettler2020,jimenez2021,cui2023,Boos2019,shi2022}.  These properties of SrCu$_2$(BO$_3$)$_2$ are well captured by the phase diagram of SS model. It is usually believed that the phase transition between the PS and N{\' e}el phase is direct~\cite{lauchli2002,corboz2013}, and the transition point could even be a deconfined quantum critical point (DQCP)~\cite{lee2019a,cui2023}. However, some recent numerical studies on SS model argued the possibility of an intermediate gapless spin liquid between PS phase and N{\' e}el phase~\cite{yang2022,keles2022,wang2022}. Since SrCu$_2$(BO$_3$)$_2$ can be fabricated quite cleanly in experiment, disorder effect is quite small which makes it promising to explore the possibility of a gapless spin liquid. It is thus also important to study some variants of SS model to find gapped spin liquids, which will guide the experimental discovery of such exotic phases. This is the motivation of this paper.

In this paper, we demonstrate the existence of a topologically ordered CSL in the highly frustrated region of SS model. Specifically we consider antiferromagnetic Heisenberg interactions in the original SS model, as well as an additional spin chirality interaction.
Using large-scale ED and DMRG calculations~\cite{white1992,schollwock2011}, we establish a global phase diagram with various conventional phases identified by local order parameters, including nonmagnetic DS, PS phase, and magnetic N{\' e}el and collinear phases.
Moreover, among the phase boundaries of these conventionally ordered phases there exists a CSL state. This CSL possesses two topologically degenerate ground states, and is characterized by modular matrix~\cite{wen1990,nayak2008} and sequential level countings in entanglement spectrum~\cite{li2008}, as the finger-prints of the $\nu=1/2$ bosonic Laughlin state processing semion anyonic statistics.

\section{Model and method}

\begin{figure}
	\centering
	\includegraphics[width=0.5\textwidth]{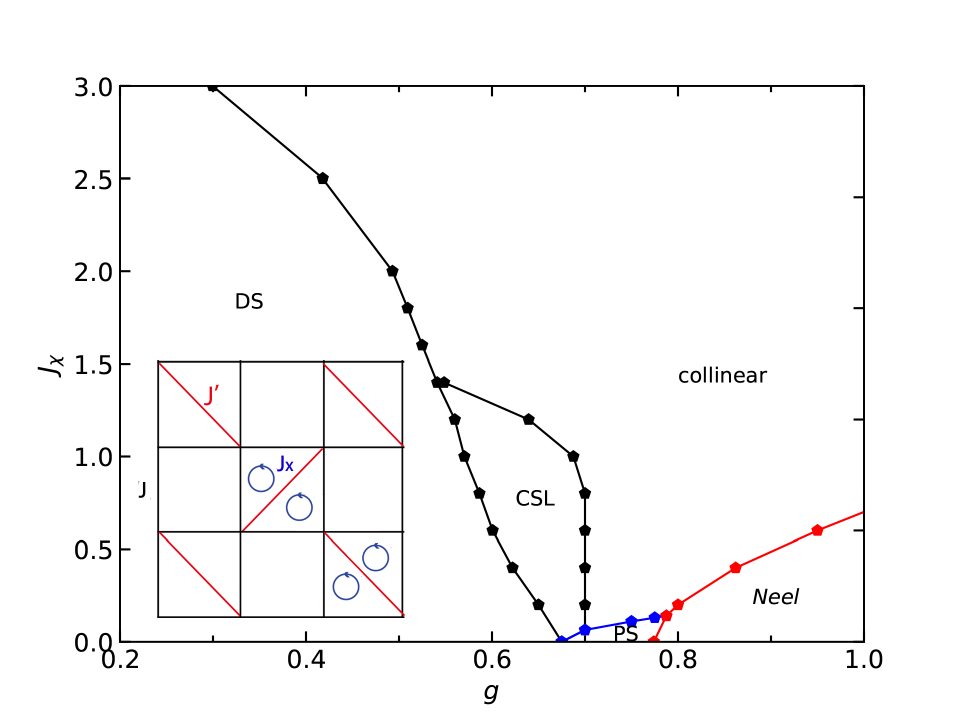}
	\caption{(Color online) 
 Quantum phase diagram versus $J_\chi-g$ of the antiferromagnetic Heisenberg model on the SS lattice.
 The phase boundaries are based from fidelity susceptibility and energy level crossing on a 36-site torus using ED method.
 Inset shows the schematic plot of SS lattice, with NN $J$ (denoted as black lines) and NNN $J^\prime$ (denoted as red lines) Heisenberg interactions, and a three-site chiral interaction (denoted as blue circles) in plaquettes with a $J^\prime$ term (see Eq.\eqref{eq:ham}). 
 }
	\label{pd}
\end{figure}

As is illustrated in the inset of Fig.~\ref{pd}, We study the SS model with an additional spin chirality interaction on a square lattice, with the Hamiltonian 
\begin{align}\label{eq:ham}
    H =& J\sum_{\left<ij\right>} \s_i\cdot\s_j + J^\prime\sum_{\left<ij\right>^\prime} \s_i\cdot\s_j \nonumber \\
    &+ J_\chi\sum_{ijk\in\bigtriangleup} \s_i\cdot(\s_j\times\s_k),
\end{align}
where antiferromagnetic Heisenberg interactions $J$ and $J^\prime$ run over all nearest-neighbor (NN) bonds $\left<ij\right>$ and specific next-nearest neighbor (NNN) bonds $\left<ij\right>^\prime$, respectively. 
The first two terms of Eq.(\ref{eq:ham}) constitute the canonical SS model which is widely accepted as the microscopic Hamiltonian describing the dominant magnetic interactions in $SrCu_2(BO_3)_2$.
The real material also features additional interactions (e.g., strong spin-orbit coupling) that has motivated the generalization of the canonical SS model to include (effective) realistic interactions. In the present study, 
we consider a three-site spin chiral interaction with strength $J_\chi$, which runs over the two triangles in each plaquette with a $J^\prime$ bond, where the sites $ijk$ are ordered in counterclockwise manner. Chiral interactions normally arise from the coupling of electrons to an external magnetic field in strong Mott insulators, particularly in lattice geometries with triangular units\cite{wen1989,baskaran1989,DSen1995,bauer2014,Samajdar2019}. It  breaks reflection (mirror-plane) and time-reversal symmetries, but their product is preserved, as are translational and rotational symmetries.  
In the following we define $g\equiv J/J^\prime$ and set $J^\prime=1$ as the energy scale.

Here we use numerically unbiased ED and DMRG method to study this model, which allows us to faithfully explore possible quantum phases and phase transitions in strongly correlated systems. Using a torus geometry in the ED calculation, we examine different system symmetries and block diagonalize the Hamiltonian in different symmetry sectors labeled by conserved quantum numbers. This allows to achieve larger system sizes in ED, and more importantly detect intriguing phase transitions using energy level crossings.
In the following, we obtain low-lying energy spectrum and label each state by quantum numbers $(S,k_x,k_y,r)$, where $S$ is total spin quantum number, $k_x$ and $k_y$ label lattice momenta in $x$ and $y$ directions respectively, and $r$ is the quantum number of the $C_4$ rotational symmetry. We can thus obtain an approximate phase diagram based on energy level crossing and fidelity susceptibility. Different order parameters are further utilized to identify the nature of quantum phases in the phase diagram. Furthermore, We find a finite region of CSL at intermediate $g$, whose topological order is elaborately characterized by both ED and DMRG calculations.
Compred with ED method, DMRG allows to study much larger system sizes in a cylinder geometry, and quantum entanglement can be straightforwardly extracted from the ground state.


\section{Phase diagram}

In this section we summarize our main results and present a global quantum phase diagram of the model in Fig.~\ref{pd}, which is based on a 36-site torus geometry using ED method. Detailed discussions of quantum phases and phase transitions will be presented in subsequent sections.

When the chiral interaction is turned off, the model returns to the well studied SS model. At small $g$ the intra-dimer $J^\prime$ term dominates and the system favors the DS phase, where each separate bond connected by the $J^\prime$ term forms a spin singlet to simultaneously minimize system energy. At large $g$ limit, the Hamiltonian reduces to the antiferromagnetic Heisenberg model on a square lattice, whose ground state possesses the conventional N{\' e}el order~\cite{white2007}. In the intermediate region, $J$ and $J^\prime$ interactions compete with each other and the SS model is highly frustrated. In this case, nonmagnetic phases with unconventional properties may also emerge. In our 36-site ED calculatoin, we find a PS phase for $g\in (0.675,0.774)$, which evolves to DS phase and N{\' e}el phase for small and large $g$ respectively. Recent calculations with careful finite-size extrapolation have also argued for the possibility of an intermediate gapless spin liquid between the PS and N{\' e}el phase \cite{yang2022}.

Next we investigate the effect of nonzero $J_\chi$ interaction. The result of our investigation is summarized in Fig.~\ref{pd}. At small to moderate values of $g$ ($\lesssim 0.5$), the ground state remains in the DS phase up to strong values of $J_\chi$, but eventually there is a transition to a collinear phase at large $J_\chi$. On the other hand, the N{\' e}el phase (at large $g$) survives for small to intermediate values of $J_\chi$, beyond which the collinear order sets in. The PS phase is the most unstable to the effects of chiral interaction and survives only for small values of $J_\chi$. It is at intermediate values of $g$ ($0.5 \lesssim g \lesssim 0.7$) that the most interesting physics emerge. 
We find clear signatures of CSL phase 
over a finite range of $J_\chi$ for these values of $g$ 
where interactions are highly frustrated. The emergence of a CSL when turning on the chiral interaction in highly frustrated region has also been observed in a number of lattice systems of Heisenberg antiferromagnet, including kagome \cite{bauer2014}, triangular \cite{gong2017} and square lattice \cite{nielsen2013}. Recent numerical studies also report the appearence of non-Abelian CSLs in higher spin systems on the square lattice~\cite{huang2022,luo2023}. Interestingly, while the CSL phase is driven by the chiral interaction, it requires the compteting Heisenberg interations for stabilization. For example, the CSL phase does not emerge in the presence of only the $J_\chi$ term in the Hamiltonian; instead, the collinear phase is realized in this limit.

\section{Conventional Orders}
\begin{figure}[t]
\centering
\includegraphics[width=0.5\textwidth]{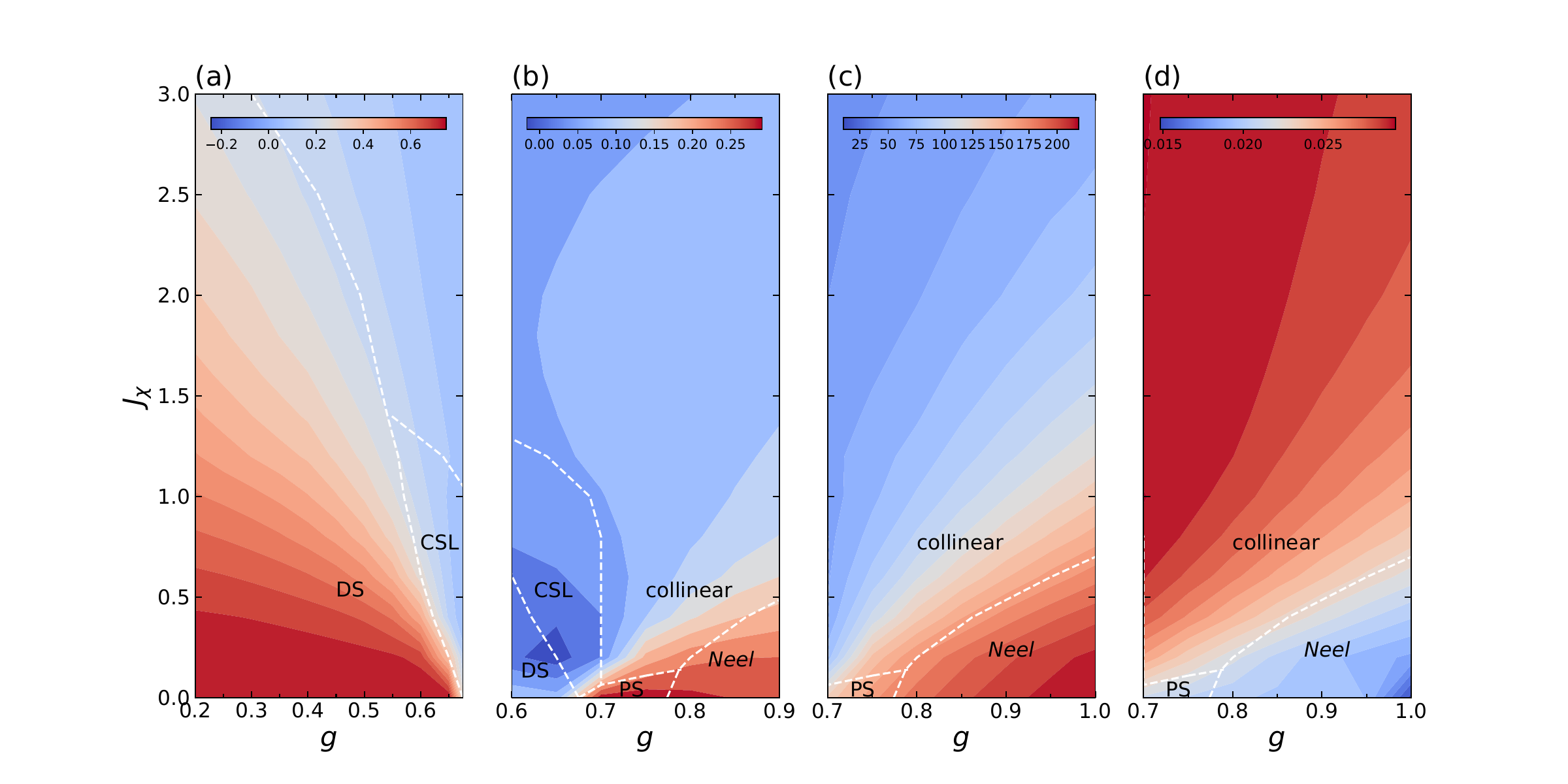}
\caption{(Color online) Profile of various order parameters measured in the quantum phase diagram. Here we study a $N=36$-site torus by ED method. (a) Dimer order parameter $O_d$, (b) plaquette order parameter $O_p$, (c) N{\' e}el order parameter $m^2(\pi,\pi)$, and (d) collinear order parameter $m^2(0,\pi)$. Dashed lines mark the same phase boundaries as in Fig.~\ref{pd}.
}
\label{order}
\end{figure}

Various conventional phases in the phase diagram can be effectively identified by suitably chosen local order parameters. First, since the dimer phase is defined by the spin singlet formed along $J'$-bonds, we define a dimer order parameter as the difference between averaged expectation value $\braket{\mathbf{S}_i\cdot\mathbf{S}_j}$ for NN $J$ bonds and the NNN $J^\prime$ bonds as
\begin{equation}
\mathbf{O}_{d}=E_{J}-E_{J^\prime}.
\end{equation}
The DS phase in the original SS model is an exact dimer phase, which exhibits a finite value of $\mathbf{O}_{d}=0.75$~\cite{corboz2013,Zhang2015}. When $J_\chi>0$, DS phase is destroyed gradually, with $\mathbf{O}_{d}$ decreasing and potentially becoming negative. As it is shown in Fig.~\ref{order}(a), $O_d$ is quite large for relatively small $g$ regime, and the region for $J_\chi=0$ with large $O_d$ coincides with the DS phase determined by other way \cite{yang2022}.


Second, we define a plaquette order parameter as
\begin{equation}
\begin{aligned}
\mathbf{O}_{p}
=&\frac{1}{2}(\mathbf{P_{\Box}}+\mathbf{P_{\Box}}^{-1})\\
=&-\frac{5}{8}+\frac{1}{4}(\mathbf{S_i}+\mathbf{S_j}+\mathbf{S_k}+\mathbf{S_l})^2+2(\mathbf{S_i}\cdot\mathbf{S_j})(\mathbf{S_k}\cdot\mathbf{S_l})\\
&+2(\mathbf{S_i}\cdot\mathbf{S_l})(\mathbf{S_j}\cdot\mathbf{S_k})-2(\mathbf{S_i}\cdot\mathbf{S_k})(\mathbf{S_j}\cdot\mathbf{S_l}),
\end{aligned}
\end{equation}
where $\mathbf{P}_{\Box}$ is a clockwise permutation operator that acts on the plaquette without a $J^\prime$ bond and $\{i,j,k,l\}$ are indicators of clockwise arrangement on the four vertices of the plaquette. $\mathbf{O}_{p}$ represents the strength of plaquette, which can be used to identify PS phase. In Fig.~\ref{order}(b), we find the plaquette order takes large value in a window for intermediate $g$. By increasing  $J_\chi$, the plaquette order monotonically decreases, which shows that the plaquette order is not favored by the spin chirality term. However, it is hard to distinguish PS and N{\'e}el phase from $\mathbf{O}_{p}$, one reason is that PS-N{\'e}el is a continuous~\cite{lee2019a} (or weakly first-order~\cite{corboz2013,zhao_symmetry-enhanced_2019}) phase transition, and another reason is that the lattice size is still too small.

Third, besides the two kinds of dimer phases, we also calculate the static structure factor to identify possible long-range magnetic orders in the phase diagram, defined as
\begin{equation}
\begin{aligned}
    m^2(k_x,k_y)=\frac{1}{N^2}\sum_{ij}e^{i\mathbf{k}\cdot (\mathbf{r}_i-\mathbf{r}_j )}\langle \mathbf{S}_i \cdot \mathbf{S}_j \rangle.
\end{aligned}
\end{equation}
A peak at $\mathbf{k}=(k_x,k_y)=(\pi,\pi)$ corresponds to antiferromagnetic N{\' e}el order, while that at $(\pi,0)$ or $(0,\pi)$ signals the formation of collinear order. In particular, $m^2(k_x,k_y)$ at $\mathbf{k}=(k_x,k_y)=(\pi,\pi)$ is also called squared magnetization. The results of $m^2(\pi,\pi)$ and $m^2(0,\pi)$ are shown in Fig.\ref{order}(c) and Fig.\ref{order}(d), respectively.

In Fig.\ref{order} (a)-(d), we find a large value of $O_p$, $m^2(\pi,\pi)$, $m^2(0,\pi)$ at PS, N{\' e}el, collinear phase, respectively. These distributions of order parameters in the phase diagram in Fig.~\ref{order} generally agree with the phase diagram presented in Fig.~\ref{pd}.

\section{Chiral spin liquid}

\begin{figure}
\centering
\includegraphics[width=0.5\textwidth]{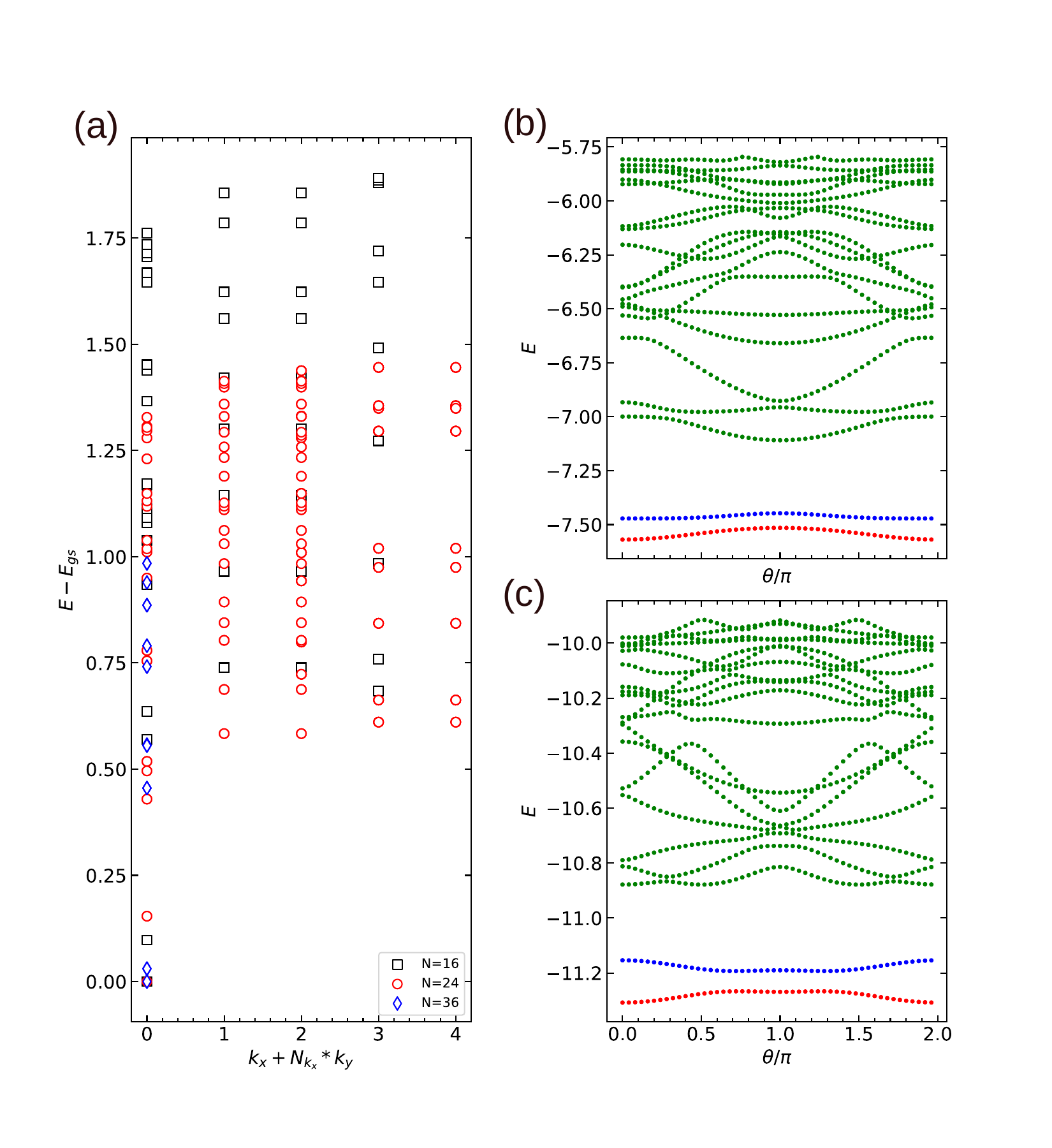}
\caption{(Color online) Energy spectrum and spectral flow of CSL phase. (a) Low-energy spectra $E_n-E_1$ versus the momentum $k_x + N_{k_x}*k_y$ in the CSL phase for lattice sizes $N=$16, 24 and 36 at ($g$,$J_{\chi}$)=(0.6,0.75), where a large gap exists between two ground states and other excited states in all studied lattice sizes. (b) and (c) are low-energy spectra versus inserted flux $\theta$ at ($g$,$J_{\chi}$)=(0.6,0.75) for $N$=16 and $N$=24, respectively, where the two topologically degenerate ground states (marked as red and blue points) remain well separated from higher energy levels (marked as green points) on flux insertion.
}
\label{spectral_flow}
\end{figure}

Among these conventional phases characterized by local order parameters, we find a finite region of CSL in the quantum phase diagram where competing interactions are highly frustrated. This spin disordered phase cannot be characterized by conventional order parameters, but by its intriguing topological order. 
Next, we will demonstrate the numerical evidence from the view points of energy spectra, entanglement spectra and modular matrix.

\subsection{Energy spectra}
Firstly, we find two-fold topologically degenerate ground states, which are well separated from bulk excitations with a large energy gap, as is shown in Fig.~\ref{spectral_flow}(a) for CSL phase of lattice sizes $N=16$, $24$ and $36$ at $(g,J_{\chi})=(0.6,0.75)$. This excitation gap remains robust when adiabatically twisting the boundary condition, i.e. the two-fold ground states never cross with the excited states. Since the two-fold ground states share the same momentum quantum numbers in these three high-symmetric lattice cluster, the energy spectrum evolves back to itself by inserting a flux quantum, as is shown in Fig.~\ref{spectral_flow}(b) and (c) for $N=16$ and $N=24$, respectively.

\subsection{Fractional statistics}
An important feature of topological ordered state is that its quasiparticle excitation takes the fractional statistics. Here we utilize the framework of modular matrices~\cite{wen1990,nayak2008}, which encode complete information of topological order, to characterize the underlying anyon statistics of the CSL. Interestingly modular matrices can be effectively obtained from entanglement measurements~\cite{zhang2012,wzhu2013}. Given the topologically degenerate ground states on a torus geometry, we construct two sets of minimally entangled states (MESs) along interwinding cuts, and the transformation between these two MES sets yields desired modular matrices.

\begin{figure}
\centering
\includegraphics[width=0.5\textwidth]{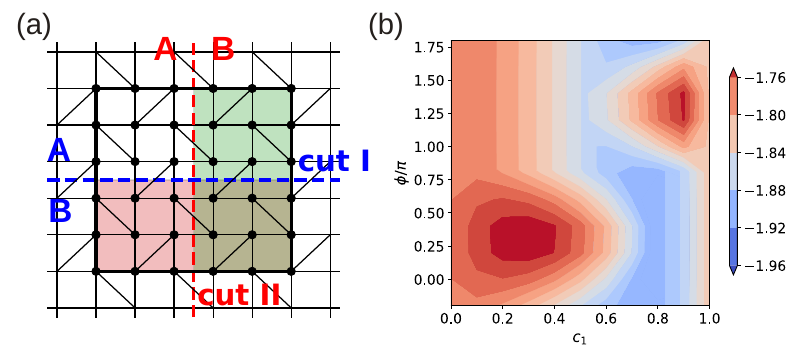}
\caption{(Color online) (a)Bipartitions of the SS model system on a $N=36$ cluster. The two ways to partition the system along the dashed lines are labeled as cut $I$ and cut $II$, respectively. (b) Profile of minus entanglement entropy (-$S$) of wave function $|\Phi_{c1,\phi}\rangle$ on the $N$=36 lattice with $g$=0.6 and $J_\chi$=0.75. These two peaks correspond to minimally entangled states along bipartition cut $I$, which are used in the calculation of modular matrix.}
\label{MES}
\end{figure}

To construct MESs, we take an arbitrary superposition of the ground states $\ket{\xi_1}$ and $\ket{\xi_2}$,
 \begin{equation}
|\Phi_{c1,\phi}\rangle = c_1|\xi_1\rangle + c_2e^{i\phi}|\xi_2\rangle
\end{equation}
where $c_1$ and $\phi$ are two independent real numbers and $c_2 = \sqrt{1-c_1^2}$. We bipartite the whole system into two parts A and B, construct the reduced density matrix of subsystem A by tracing out degrees of freedom in subsystem B as $\rho_A$=$\mathrm{Tr}_B|\Phi_{c1,\phi}\rangle\langle\Phi_{c1,\phi}|$, such that the entanglement entropy for this cut is obtained by $S=-\log\mathrm{Tr}\rho_A^2$. By evaluating $S$ for each linear combination of the ground states, we identify ($c_1$,$\phi$) pairs corresponding to states with minimal entanglement entropy, i.e. MESs.
As it is shown in Fig.\ref{MES}(a), we use two ways to bipartite the system along noncontractable cuts $I$ and $II$, each yielding one set of MESs. In Fig.~\ref{MES}(b), we draw the profile of -$S$ as a function of ($c_1$,$\phi$) in the contour plot for 36-site lattice, so that the peaks represent the minimums of $S$.  In this way, we identify two peaks in ($c_1$,$\phi$) parameter space, corresponding to two distinct MESs,
 \begin{equation}
 \begin{aligned}
|\Xi_1^I\rangle &= 0.897|\xi_1\rangle + 0.442e^{i1.374\pi}|\xi_2\rangle \\
|\Xi_2^I\rangle &= 0.295|\xi_1\rangle + 0.955e^{i0.371\pi}|\xi_2\rangle
\end{aligned}
\end{equation}
where the label $I$ means we bipartite the system along the horizontal line (cut $I$). We find the relative phase difference between these two MESs is $\phi(1)-\phi(2)=\pi$, and consequently the two MESs are approximately orthogonal to each other for this finite size system: $|\langle\Xi_1^I |\Xi_2^I \rangle| \approx 0.169$. Due to the $\pi/2$ rotation symmetry in the system, the MESs along the vertical line (cut II) $|\Xi_i^{II}\rangle$  are related to  $|\Xi_i^{I}\rangle$ as $|\Xi_i^{II}\rangle=R_{\pi/2}|\Xi_i^{I}\rangle$, where $R_{\pi/2}$ rotates a state anticlockwisely by $\pi/2$.

In our calculation, we identify that
\begin{equation}
\begin{aligned}
R_{\pi/2}|\xi_1\rangle &=  |\xi_1\rangle \\
R_{\pi/2}|\xi_2\rangle &=  -|\xi_2\rangle
\end{aligned}
\end{equation}
and then we have the second set of MESs,
\begin{equation}
\begin{aligned}
|\Xi_1^{II}\rangle &= 0.897|\xi_1\rangle - 0.442e^{i1.374\pi}|\xi_2\rangle \\
|\Xi_2^{II}\rangle &= 0.295|\xi_1\rangle - 0.955^{i0.371\pi}|\xi_2\rangle
\end{aligned}
\end{equation}
Finally we get the modular matrix by $S=\langle\Xi^{II} |\Xi^I \rangle$ as
\renewcommand\arraystretch{2}
\begin{equation}
S\approx0.7176
\begin{pmatrix}
0.8490 &  0.9713\\
0.9713 & -1.1510

\end{pmatrix}
\end{equation}
which is close to the analytic prediction for the bosonic $\nu=1/2$ Laughlin state~\cite{dong2008,fendley2007,rowell2009}:

\begin{equation}
S=\frac{1}{\sqrt{2}}
\begin{pmatrix}
1 &  1\\
1 & -1
\end{pmatrix}.
\end{equation}
This directly shows the semion statistics emergent in the CSL. 

\begin{figure}
\centering
\includegraphics[width=0.5\textwidth]{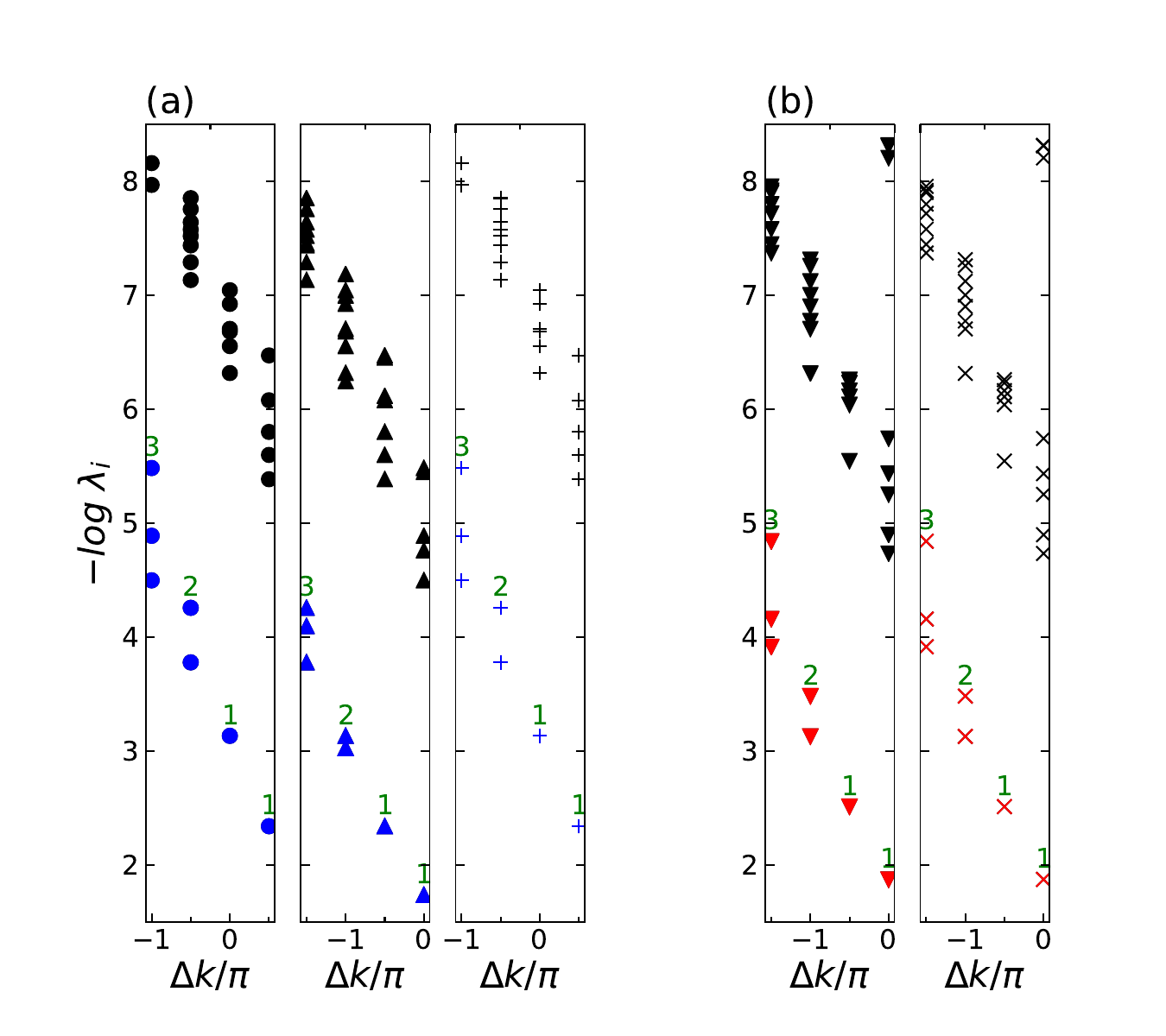}
\caption{(Color online) Momentum resolved entanglement spectra of the two ground states at $(g,J_\chi) = (0.6,0.75)$ on the 8$\times$16 cylinder, where $\Delta k$ means relative momentum in transverse direction of the cylinder and $\lambda_i$ represents the eigenvalues of reduced density matrix $\rho_A$.} Here we set $\Delta k=0$ for largest $\lambda_i$. (a) and (b) corresponds to the ground state at vacuum and semion sector, respectively. These counting structures at different $S^z$ sectors match the tower of states of $SU(2)_1$ Wess-Zumino-Witten theory. 

\label{es}
\end{figure}

\subsection{Entanglement spectra}
To establish the existence of CSL conclusively, we have studied larger systems, up to $L_y\times L_x = 8\times 16$ with cylindrical lattice using DMRG for a representative set of parameters for which the ground state is in the CSL phase. Various entanglement information of the ground state can be straightforwardly extracted in this method. To reveal the underlying topological order, we calculate the entanglement spectrum~\cite{li2008} from the entanglement Hamiltonian $-\log\rho_A$, whose low-lying spectrum $\{-\log\lambda_i\}$ harbors characteristic edge countings for topologically ordered phases.

In the DMRG optimization we get one of the ground states directly and the other one by the removal of a single site at each edge. We keep up to 1200 $U(1)$ states to obtain accurate results with the truncation error less than $10^{-6}$. The entanglement spectra of the ground states are presented in Fig.~\ref{es}. 
The degeneracy pattern in low-lying entanglement spectrum $\{1, 1, 2, 3, 5, . . . \}$ follows the $U(1)$ decomposition of the $SU(2)_1$ Wess-Zumino-Witten CFT theory~\cite{francesco1997} exactly, establishing the same topological order as $\nu=1/2$ bosonic Laughlin state. Intuitively, this counting structure can also be simply obtained by generalized Pauli principle~\cite{bernevig2008}, which states no more than 1 particle in two consecutive orbitals in this case.


\begin{figure}
\centering
\includegraphics[width=0.5\textwidth]{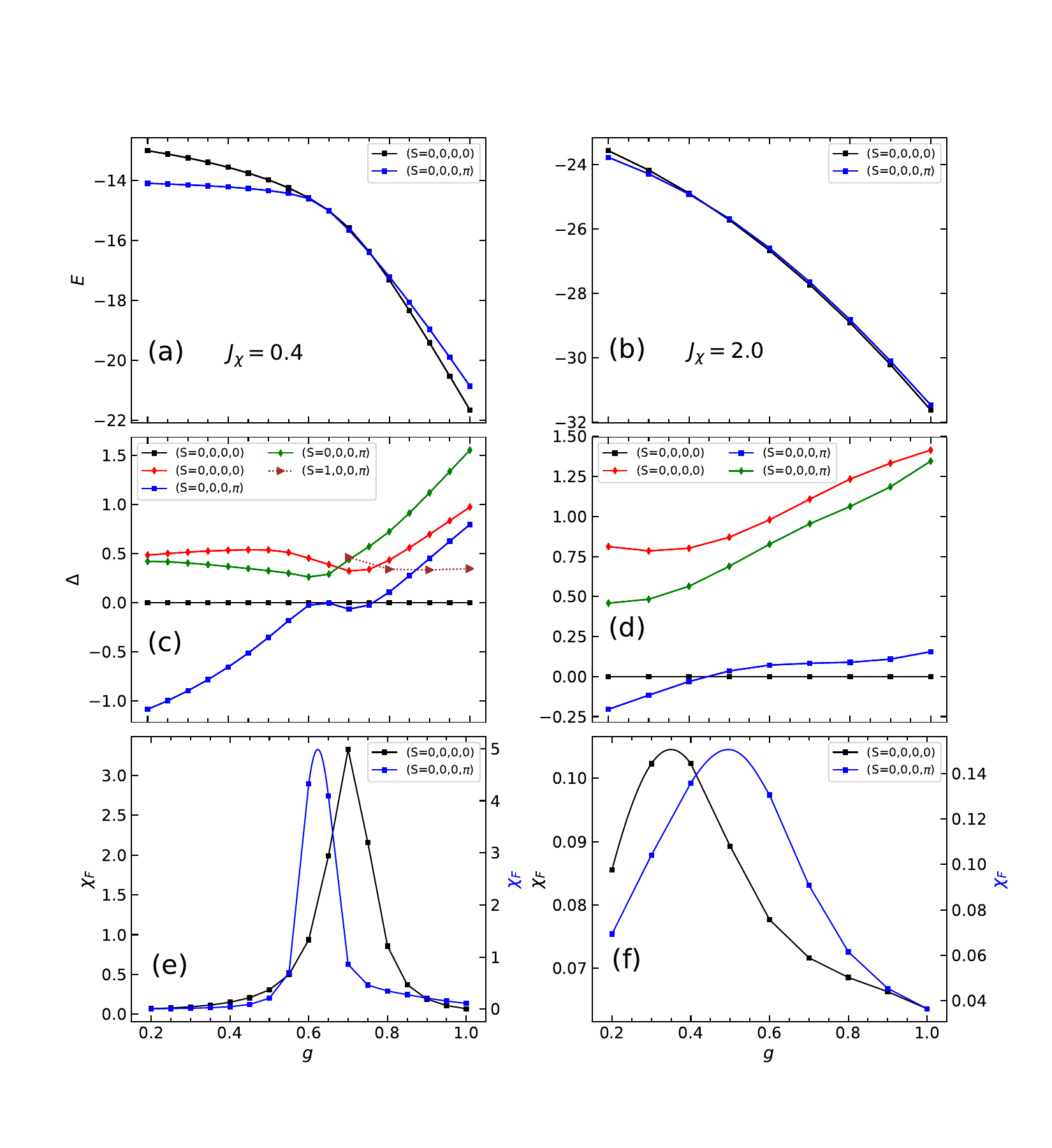}
\caption{(Color online) Level spectroscopy at $J_\chi=0.4$ and 2.0 in (a,b) respectively by varying $J_2$ for N=36 periodic cluster. Two lowest singlets are displayed for both the (S=0,0,0,0) sector (in black squares and red pentagons) and (S=0,0,0,$\pi$) sector(in blue squares and green pentagons). Characteristic magnetic excitation of the N{\'e}el phase are also shown, i.e. Anderson Tower excitation in terms of the (S=1,0,0,$\pi$) triplet (in purple triangles). Energy gaps defined with respect to the lowest (S=0,0,0,0) singlet are shown in (c,d) for $J_{\chi}=0.4$ and $2.0$ respectively. For the lowest (S=0,0,0,0) singlet and the lowest (S=0,0,0,$\pi$) singlet state, the corresponding FS are shown in (e,f).
}
\label{level0}
\end{figure}

\section{Phase boundaries}

In this section we 
determine the phase boundaries between the various ground state phases in Fig.\ref{pd}, based on energy level crossing and fidelity susceptibility numerically using ED method on a 36-site torus. In order to get the phase boundaries, we fix $J_\chi$ and change $g$. As examples, the results of $J_\chi=0.4$ and $J_\chi=2.0$ are shown in Fig.\ref{level0}. All states $\vert\phi\rangle$ are labeled by quantum numbers $(S,k_x,k_y,r)$. In our results, the low-lying states in the level spectroscopy are mainly located in $r=0$ and $r=\pi$ sectors, where $r=0$ means the state satisfy $R_{\pi/2}\vert\phi\rangle=\vert\phi\rangle$, while $r=\pi$ means 
$R_{\pi/2}\vert\phi\rangle=-\vert\phi\rangle$.

Fidelity susceptibility (FS) we use here is defined as 
\begin{equation}
\chi_F(g)\equiv {\lim_{ \delta g \to 0}} \frac{-2\ln F }{ (\delta g)^2}=-\frac{\partial^2F}{\partial(\delta g)^2},
\end{equation}
where $F=\ln \vert \langle\phi(g+\delta g)|\phi(g)\rangle\vert$ is the definition of fidelity and $\phi(g)$ is the “ground state” of a given singlet sector at parameter $g$. We compute FS in both (S=0,0,0,0) and (S=0,0,0,$\pi$) sectors, which are shown in Fig.~\ref{level0}(e) and Fig.~\ref{level0}(f). When quantum phase transition occurs, $\phi(g)$ undergoes a significant change at a certain point, the fidelity $F$ deviates from 1, leading to a peak on FS curve.

FS approach is effective in detecting quantum phase transitions \cite{PhysRevA.81.064301,PhysRevB.84.174426,pub.1062937215}. With this approach, we obtain the DS-CSL, DS-collinear and CSL-collinear phase boundaries i.e. the black curves in Fig.~\ref{pd}. However, for some certain types of quantum phase transitions, such as the PS-N{\'e}el transition in the original SS model ($J_\chi=0$), there are no peaks on fidelity curves and the FS approach fails. 

For this reason, we use excited energy level crossings to get the PS-N{\'e}el, PS-collinear and N{\'e}el-collinear phase boundaries, i.e. the red and blue curves in Fig.~\ref{pd}. From this point of view, quantum phases have their own characteristic energy spectra, which is usually called tower of states (TOS), and the quantum phase transitions are caused by the reconstruction of TOS. During the reconstruction, the crossing point of low-lying excited energy levels provides an approximate quantum critical point at a finite size. This method has been widely applied to various different systems~\cite{okamoto1992,eggert1996a,wang2018,ferrari2020,nomura2021,sandvik2010a,suwa2015,suwa2016,yang2022,wang2022}.

In the study of the square lattice and the SS lattice, both FS and level crossing methods have also been successfully applied in Ref. \cite{wang2018,yang2022,wang2022}. Below we will have a detailed discussion on Fig.~\ref{level0}.

We first review the phase transitions in the original SS model at $J_\chi=0$. The energy evolution of the ground states on increasing NN Heisenberg interaction $J$ (or equivalently $g$) is easy to be understood. The two lowest singlet(S=0) energy levels 
cross with each other and change their order at $g\approx0.675$. For small $g\lesssim0.675$ the global ground state 
has a relatively fixed energy on varying $g$, which is a typical character of the DS phase. In this case the ground state is mainly a product state of spin singlets, with each bond connected by the $J^\prime$ term. As a result, the typical ground state energy is thus $-0.75N_s$.
For $g\gtrsim0.675$ the global ground state instead changes to other phases, namely PS phase and N{\' e}el phase in this region~\cite{lauchli2002,corboz2013,Zhang2015,lee2019a}. 
(It is argued recently that there may also exist an intermediate gapless spin liquid phase between PS and N{\' e}el phase~\cite{yang2022,wang2022}). 
Energy level crossing may be the most effective approach to distinguish PS and N{\' e}el phase, since the low-lying states of N{\' e}el phase is the famous Anderson TOS, while the ground states of  PS have a 2-fold degeneracy~\cite{lauchli2002,mambrini2006}. The lowest excitation of N{\' e}el phase is a triplet state, while it is a singlet state for collinear phase.
We can use the singlet(S=0)-triplet(S=1) crossing point as the boundary of PS phase in small cluster. 
This point is also in great agreement with the iPEPS~\cite{corboz2013} and iDMRG~\cite{lee2019a} results~\cite{yang2022}.

However, as it is shown in Fig.~\ref{level0}(a) and Fig.~\ref{level0}(c), the energy spectra at $J_\chi=0.4$ is markedly different. For $g\lesssim 0.625$, the ground state is still a DS phase, and the peak at $g\approx0.625$ of FS curve in (S=0,0,0,$\pi$) sector (see Fig.\ref{level0}(e)) forms a boundary of DS. When $g>0.625$ there is a finite range of $g$ over which the two ground states in the (S=0,0,0,0) and (S=0,0,0,$\pi$) sectors are nearly degenerated, which turn out to be the topologically degenerated ground states of CSL phase. In Fig.~\ref{level0}(e), the FS curve of the ground state in (S=0,0,0,0) sector indicates a phase transition at $g=0.7$, which becomes a phase boundary of CSL phase. Further increasing $g$ leads to two magnetic orders, i.e. collinear order and N{\' e}el order. They are separated by the singlet-triplet level crossing point at $g=0.862$, which are displayed in Fig.~\ref{level0}(c). Since the low-lying states of N{\' e}el order form Anderson TOS, the lowest excitation is a triplet state in (S=0,0,0,$\pi$) sector, while for collinear order, the lowest excitation is one of the 2-fold degenerated ground states, which is a singlet state in (S=0,0,0,$\pi$) sector.

When $J_{\chi}$ is turned to be 2.0, the ground state is still DS state living in (S=0,0,0,$\pi$) sector while $g$ is small. With the increase of $g$, the state in (S=0,0,0,$\pi$) sector has a FS peak at $g\approx0.5$ as shown in Fig.\ref{level0}(f). At the same time, there is a crossing point between (S=0,0,0,0) and (S=0,0,0,$\pi$) curves at $g=0.45$ (see Fig.\ref{level0}(d)). Both points can be regarded as the DS-collinear phase boundary, but they are not consistent due to the finite size effect. In order to keep consistent with the case of $J_\chi=0.4$, we select the FS peak at $g\approx0.5$ as the phase boundary. Similar as Fig.\ref{level0}(e), in Fig.\ref{level0}(f) both (S=0,0,0,0) sector and (S=0,0,0,$\pi$) sector have a FS peak. However, the phase boundaries presented in these two figures are completely different. In Fig.\ref{level0}(f), the FS peak of (S=0,0,0,0) curve is located at $g=0.35$, indicating a phase transition of the excited state of DS phase, but only when $g>0.5$ the corresponding states of (S=0,0,0,0) and (S=0,0,0,$\pi$) sectors will form the 2-fold degeneracy of the collinear phase.

\begin{figure}[t]
\centering
\includegraphics[width=0.5\textwidth]{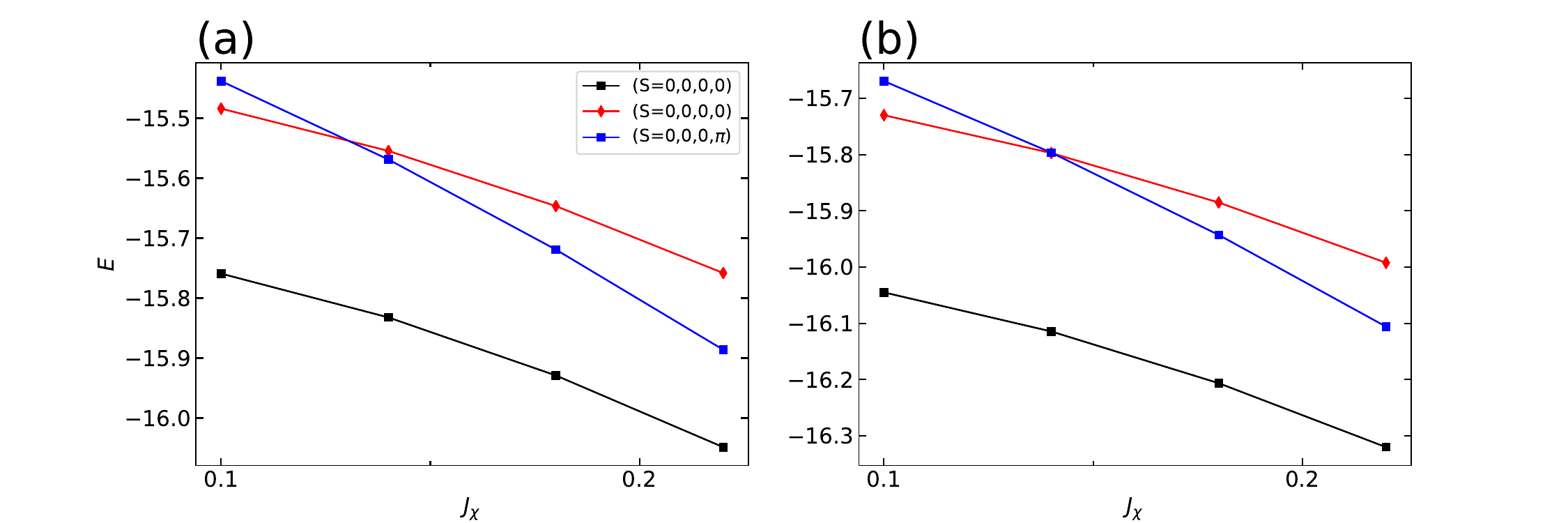}
	\caption{(Color online)  Level spectroscopy at $g=0.775$ and 0.7875 in (a,b) respectively by varying $J_\chi$ for N=36 periodic cluster. Two lowest singlets in the (S=0,0,0,0) sector are displayed in black squares and red pentagons respectively and the lowest singlet in the (S=0,0,0,$\pi$) sector is displayed in blue squares. 
	}
	\label{level1}
\end{figure}

In order to pin down the phase boundary of PS phase we fix $g$ and instead change $J_\chi$. 
In Fig.~\ref{level1}(a) we show the evolution of three low lying energy levels on increasing the chiral interaction at $g=0.775$. At small $J_\chi$ the system is in PS phase and thus lowest two energy levels remain in (S=0,0,0,0) sector. For $J_\chi\gtrsim0.13$ an excited level crossing occurs, indicating a transition to the collinear phase, where the lowest two energy remain in (S=0,0,0,0) and (S=0,0,0,$\pi$) sector, respectively. We also observe a similar behavior at $g=0.7875$, where the excited level crossing occurs at $J_\chi\approx0.14$.

Based on above discussion, we establish the ground state phase diagram as shown in Fig.~\ref{pd}, which is further corroborated by the results of order parameters in Fig.~\ref{order}.

\section{Summary and discussion}

In summary, we study the antiferromagnetic SS model with additional three-spin chiral interaction using numerically unbiased ED and DMRG calulations. Using local order parameters, various conventional phases have been identified, including magnetic N{\' e}el phase and collinear phase, and nonmagnetic DS and PS phase. PS phase emerges when SS model is highly frustrated, and we show that a topologically ordered CSL can emerge by adding a small chirality interaction to this phase, which is conclusively identified as $\nu=1/2$ Laughlin state by characteristic modular matrix and entanglement spectrum. In addition, the phase transitions in the global phase diagram are determined by using energy level crossing and fidelity susceptibility, which shows a finite region of CSL phase. Since quantum material SrCu$_2$(BO$_3$)$_2$ is highly related to the SS model, we expect the observation of CSL under certain circumstances in experiment. To be specific, the CSL exhibits electromagnetic signatures to facilitate experimental detection even within a Mott insulator regime \cite{Saikat2023} , i.e. the electrical charge and orbital electrical current associated with a spinon excitation and  a nonvanishing optical response.



\textit{Acknowledgement.---}
W.W.L. and W.Z. were supported by ``Pioneer" and ''Leading Goose" R\&D Program of Zhejiang (2022SDXHDX0005), National Natural Science Foundation of China (No.~92165102, 11974288) P.S. acknowledges financial support from the Ministry of Education, Singapore through grant no. RG 159/19.

$*$ These authors contributed equally to this work.

\bibliography{SSchiral}
\bibliographystyle{apsrev4-1}

\end{document}